\documentclass[iop,apj]{emulateapj}

\usepackage{natbib}
\usepackage{graphicx}	
\usepackage{amsmath}	
\usepackage{amssymb}	
\usepackage{units}
\usepackage{enumitem}
\usepackage{bm}
\usepackage{color}
\usepackage[breaklinks,colorlinks,citecolor=blue,linkcolor=red]{hyperref} 

\def\msun{{\rm\,M_\odot}}

\def\msun{{\rm\,M_\odot}}

\newcommand{\kms}{\, {\rm km\, s}^{-1}}

\newcommand{\be}{\begin{equation}}
\newcommand{\ee}{\end{equation}}

\def\h2{${\rm\,H_2}$}

\newcommand{\beq}{\begin{equation}}
\newcommand{\beqa}{\begin{eqnarray}}
		 \newcommand{\eeq}{\end{equation}}
\newcommand{\eeqa}{\end{eqnarray}}
\newcommand{\mts}[2]{\textcolor{black}{{ #1}} }
\definecolor{orange}{rgb}{1, 0.5, 0}
\newcommand{\jds}[2]{\textcolor{black}{{ #1}} }

\begin{document}

\title{A Statistical Detection of Wide Binary Systems in the Ultra-Faint Dwarf Galaxy Reticulum II}
\author{Mohammadtaher Safarzadeh\altaffilmark{1,2}, Joshua D. Simon\altaffilmark{3}, Abraham Loeb\altaffilmark{1}}
\affil{$^1$Center for Astrophysics | Harvard \& Smithsonian, 60 Garden Street, Cambridge, MA; \href{mailto:msafarzadeh@cfa.harvard.edu}{msafarzadeh@cfa.harvard.edu}}
\affil{$^2$Gravitational Astrophysics Laboratory, NASA Goddard Space Flight Center, Greenbelt, MD 20771, USA}
\affil{$^3$Observatories of the Carnegie Institution for Science, 813 Santa Barbara Street, Pasadena, CA 91101, USA}

\begin{abstract}
Binary stars can inflate the observed velocity dispersion of stars in dark matter dominated systems such as ultra-faint dwarf galaxies (UFDs). However, the population of binaries in UFDs is poorly constrained by observations, with preferred binary fractions for individual galaxies ranging from a few percent to nearly unity.  Searching for wide binaries through nearest neighbor (NN) statistics (or the two-point correlation function) has been suggested in the literature, and we apply this method for the first time to detect wide binaries in a UFD. By analyzing the positions of stars in Reticulum~II (Ret~II) from Hubble Space Telescope images, we search for angularly resolved wide binaries in Ret~II.  We find that the distribution of their NN distances shows an enhancement at projected separations of $\lesssim1\arcsec$~relative to a model containing no binaries. We show that such an enhancement can be explained by a wide~binary fraction of $f_b\approx0.007^{+0.008}_{-0.003}$ at separations of more than 3000~AU.  Under the assumption that the binary separation distribution is similar to that in the Milky Way, the total binary fraction in Ret~II may be on the order of 50\%.~We also use the observed magnitude distribution of stars in Ret~II to constrain the initial mass function over the mass range $0.34-0.78~\msun$, finding that a shallow power-law slope of 
$1.01 \le \alpha \le 1.15$~matches the data.
\end{abstract}


\section{Introduction}
\label{sec:intro}

Ultra-faint dwarf galaxies (UFDs) are the lowest luminosity galaxies known, with $L < 10^5 L_{\odot}$.
UFDs are dark matter-dominated systems with ancient stellar populations \citep[e.g.,][]{SG2007ApJ,Geha2009,Brown2014,Li2018CarII,Simon2019ARAA,Sacchi2021}, 
which have been discovered at a rapid rate over the past 15 years by the Dark Energy Survey and other wide-field imaging surveys \citep[e.g.,][]{Willman2005UMaI,zucker2006CVn,Belokurov2006BooI,Bechtol2015ApJ,Koposov2015ApJ}. 
Being relics of the early universe, UFDs have challenged various theories of dark matter, from the standard collisionless dark matter picture \citep{Nadler2021PhRvL}, to ultra-light (fuzzy) dark matter \citep{Calabrese2016,SS2020ApJ,Burkert2020ApJ,Hayashi2021} and MOND \citep{MW2010ApJ,SL_MOND}.

The inferred dark matter masses and densities of dwarf galaxies are obtained through measurements of the line of sight velocity dispersion \citep[e.g.,][]{Aaronson1983,Kleyna2005,SG2007ApJ,Wolf2010}.
However, these measurements are in principle susceptible to the presence of binary stars. 
For binary stellar systems in UFDs, the primary mass has an upper limit of $0.8~M_{\odot}$ (because more massive stars have evolved into compact objects), while a typical companion mass is $\approx 0.3-0.5~M_{\odot}$ for an M dwarf and $\approx0.6~M_{\odot}$ for a white dwarf. \mts{A binary system of~$0.8$ and $0.5~M_{\odot}~$ stars}, with a period of 100 years will have a circular velocity on the order of a few $\kms$.

Such binaries can inflate the inferred velocity dispersion of the galaxy \citep[][although see \citealt{Minor2010}]{McConnachie:2010,Simon2011ApJSegI,Kirby2013ApJSeg2,Kirby2017ApJ,Ji2016BooII,Spencer2018AJ,Simon2019ARAA}. The recent determination that the close binary fraction (at separations less than 10~AU) in the Milky Way varies inversely with metallicity, reaching 50\%\ at $\textrm{[Fe/H]} = -3$ \citep{Badenes2018,Moe2019ApJ}, increases the importance of this issue for very metal-poor systems such as UFDs.  \jds{On the other hand, the wide binary fraction (separation $>1000$~AU) may decrease toward lower metallicity \citep[][although see \citealt{El-BadryRix19}]{Hwang21}, but no measurements are available for stars as metal-poor as those in UFDs.}~

Many recent studies attempt to minimize the effect of close binaries by identifying and removing them from kinematic samples using velocity measurements with a time baseline of $\sim1$~yr \citep[e.g.,][]{Fu2019,Jenkins2021,Buttry2021}.  However, somewhat wider binaries may evade detection if their velocity changes by less than $2-3~\kms$ in this time interval.  An alternative approach is to trace the population of binary stars through the distribution of the projected separations between the stars \citep{Penarrubia:2016,Kervick:2021}.  This idea has been explored conceptually, but not examined with real data as it requires  
high quality imaging with spatial resolution comparable to the expected separations of wide binaries.  The angular separations, in turn, depend on the distance to the UFD and the intrinsic distribution of projected binary separations. Moreover, in order to obtain a reliable statistical inference of the binary fraction, the imaging must be deep enough or the galaxy must be luminous enough to make a large sample of member stars available.  Given the combined considerations of distance, luminosity, and imaging coverage, we identify Reticulum~II (Ret~II) as an intriguing target for detecting wide binaries.  Ret~II is located at a distance of \jds{$31.4\pm1.4$~kpc}~\citep{MutluPakdil2018ApJ}, and recent Hubble Space Telescope (HST) imaging of the galaxy provides a sample of $\sim2600$ member stars.  At the distance of Ret~II, a binary system with a projected separation of 1 pc will have an angular separation of 6\farcs7.  The HST diffraction limit at optical wavelengths of $\sim0.1\arcsec$ corresponds to a physical scale of 0.015~pc (3000~AU).  \mts{With the observed separation distribution of wide binaries (spanning from $\sim10 - 10^{5}$~AU) in the solar neighborhood \citep{El-Badry2021MNRAS}, $\sim3\%$ of binary pairs should be resolvable at the distance of Ret~II.}

For the first time, we analyze the statistics of nearest neighbors in Ret~II in search of enhancements at close separations that could be explained by the presence of the binaries. 
The structure of this paper is as follows: in \S2, we discuss Ret~II and the data used for our analysis. In \S3, we explain how we model the distribution of the stars as an ellipsoid and how magnitudes are assigned to the stars. In \S4, we discuss our result and constraints on the derived binary fraction of Ret~II based on fitting the NN distribution, and we investigate the initial mass function in Ret~II using the observed magnitude distribution.

\section{TARGET AND OBSERVATIONS}\label{sec:data}

\subsection{Properties of Reticulum~II}
Ret~II is a typical UFD, with an absolute magnitude of $M_{V} = -3.1$ ($L = 1500~L_{\odot}$) and a half-light radius of 58~pc \citep{MutluPakdil2018ApJ}.  The important physical parameters for our analysis are its angular size and ellipticity, both of which affect the projected separations between random (physically unassociated) pairs of member stars.  Published values for these properties are: $R_{h} = 6\farcm0 \pm 0\farcm6$, $\epsilon = 0.6^{+0.1}_{-0.2}$ \citep{Bechtol2015ApJ}, $R_{h} = 3\farcm6^{+0\farcm2}_{-0\farcm1}$, $\epsilon = 0.59^{+0.02}_{-0.03}$ \citep{Koposov2015ApJ}, $R_{h} = 5\farcm4 \pm 0\farcm2$, $\epsilon = 0.56 \pm 0.03$ \citep{RMunoz2018}, $R_{h} = 6\farcm3 \pm 0\farcm4$, $\epsilon = 0.6 \pm 0.1$ \citep{MutluPakdil2018ApJ}, and
$R_{h} = 6\farcm6 \pm 0\farcm2$, $\epsilon = 0.60 \pm 0.01$ \citep{Moskowitz2020}.  The latter three studies use deeper and more uniform photometry than the discovery papers, but the results across all five structural analyses are in reasonable agreement apart from the small size determined by \citet{Koposov2015ApJ}.  To ensure consistency with our modeling procedures described in \S3, we perform our own fits to both the DES DR1 \citep{desdr1} data set employed by \citet{RMunoz2018} and \citet{Moskowitz2020} and the Megacam photometry from \citet{MutluPakdil2018ApJ}.  Following the methodology described by \citet{DrlicaWagner2020} and \citet{Simon:2021} and fitting with an exponential profile, we find $R_{h} = 5\farcm9^{+0\farcm4}_{-0\farcm3}$, $\epsilon = 0.61 \pm 0.03$ with the DES data and $R_{h} = 6\farcm2 \pm 0\farcm3$, $\epsilon = 0.60 \pm 0.03$ with the Megacam data.  As \jds{representative measurements from the deepest available data, we adopt the \citet{MutluPakdil2018ApJ} values of $R_{h} = 6\farcm3$ and (when needed) $\epsilon = 0.6$}~ for the remainder of this paper.

\subsection{HST Imaging}
Reticulum~II was observed with the Wide Field Channel of the Advanced Camera for Surveys (ACS; \citeauthor{Ford:2003}) on the Hubble Space Telescope (HST) for program GO-14766 (PI: Simon).  The data consist of 12 image tiles, each of which was observed for one orbit that was split between the F606W and F814W filters.  Data reduction and photometry followed the procedures described by \citet{Brown2014} and \citet{Simon:2021}, using DAOPHOT-II \citep{Stetson:1987} to carry out point-spread function fitting in the STMAG system.  Further details about the data and processing will be provided in a forthcoming paper analyzing the star formation history of Ret~II with the same data set (J.~D.~Simon et al., in prep.).

We select Ret~II member stars in the ACS color-magnitude diagram (CMD) based on a PARSEC isochrone track \citep{Bressan:2012,Chen:2014,Marigo:2017} with an age of 13 Gyr (similar to previously studied UFDs; \citealt{Brown2014}) and a metallicity of $\textrm{[Fe/H]} = -2.65$ \citep{Simon2015ApJRetII}.  We use a selection window in color space that extends 0.15~mag bluer than the isochrone, except for the magnitude range near the main sequence turnoff, where the window is expanded to 0.40~mag to include the blue stragglers.  To the red side, the selection window is 0.15~mag for stars brighter than $m_{814} = 22$, and gradually widens to 0.45~mag at $m_{814} = 26.5$.  We limit the member sample to magnitudes brighter than $m_{814} = 26.5$ to avoid the much greater levels of contamination and incompleteness that set in at fainter magnitudes.

\begin{figure*}
\epsscale{0.80}
\plotone{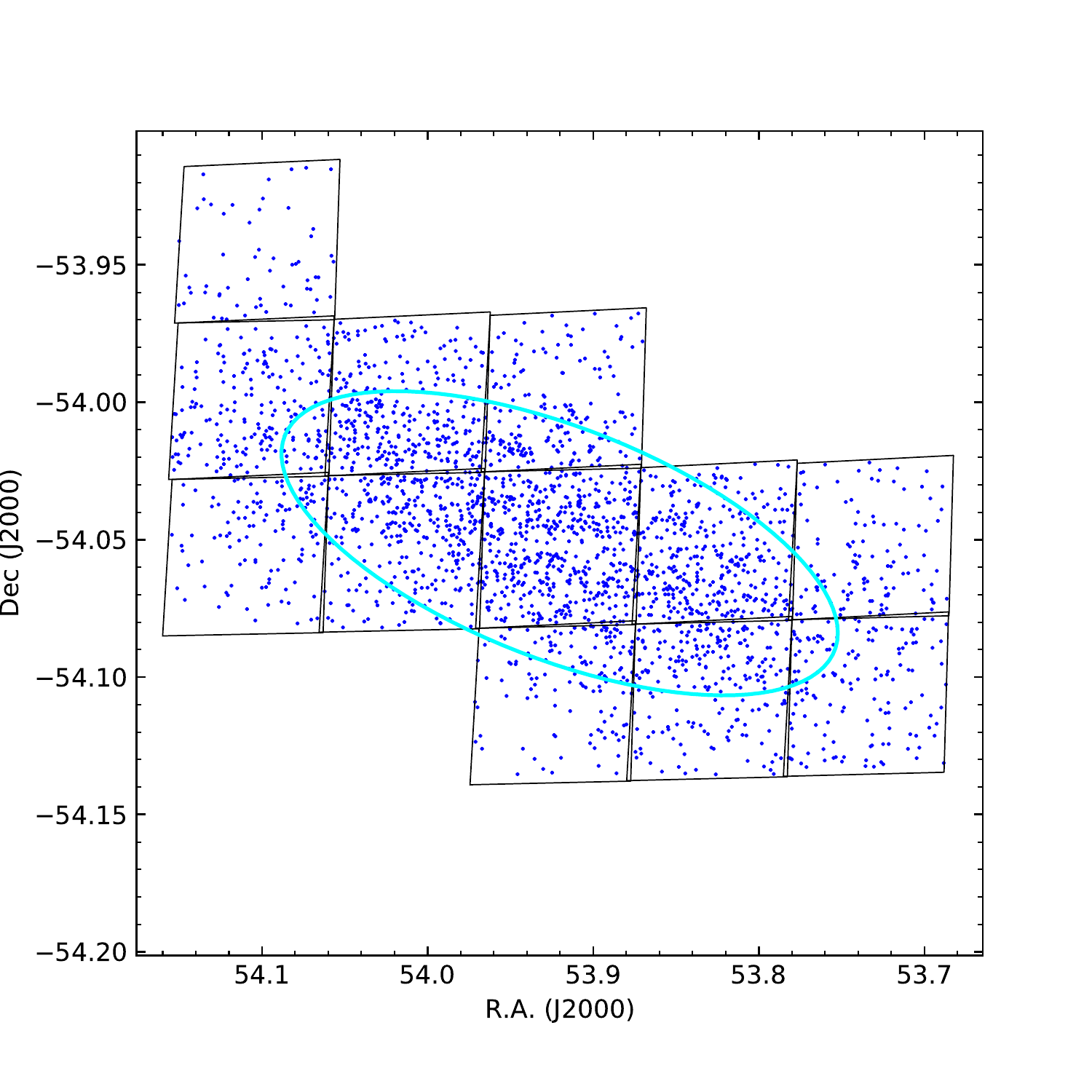}
\caption{\jds{Spatial distribution of Ret~II stars in the ACS data set.  The 12 ACS pointings are outlined in black, and an ellipse marking the half-light radius of Ret~II is shown in cyan.}~}
\label{fig:spatial_coverage}
\end{figure*}

We remove sources with bad DAOPHOT photometry flags or spatial profiles that are not consistent with a point source.  However, stars whose fluxes could be contaminated by neighboring objects are not removed from the sample, in order to avoid discarding true binaries.  For the goals of this study, small photometric errors that may result from stars with close and/or bright neighbors are unimportant.  In order to avoid artifacts from bright, saturated stars, we also eliminate detected objects that are within 5\arcsec\ of $G < 11$ stars in the Gaia eDR3 catalog \citep{GaiaCollaboration:2016,GaiaCollaboration:2021}, within 3\arcsec\ of $11 \le G < 15$ stars, or within 1\arcsec\ of $15 \le G < 17$ stars.  Finally, we check for and remove sources with multiple slightly offset detections (which result from astrometric distortions in ACS) in the small overlap regions between tiles.  After this cleaning, the total number of presumed Ret~II member stars is 2587.  In Figure~\ref{fig:spatial_coverage}, we display the spatial distribution of this sample of stars with the ACS tiling pattern overlaid.

\subsection{Distribution of Nearest Neighbor Distances}

\begin{figure*}
\hspace{-0.2in}
\centering
\epsscale{1.17}
\plottwo{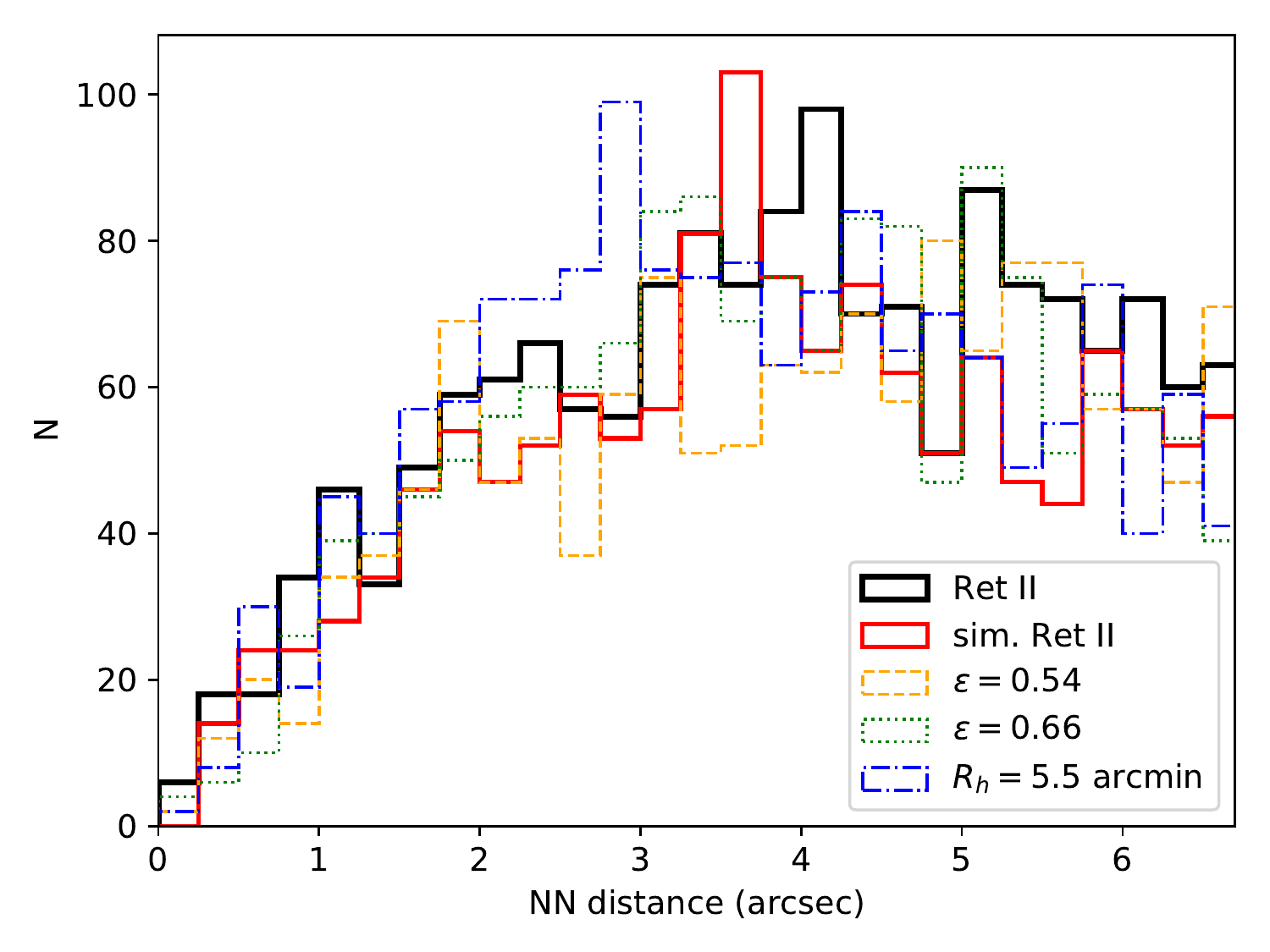}{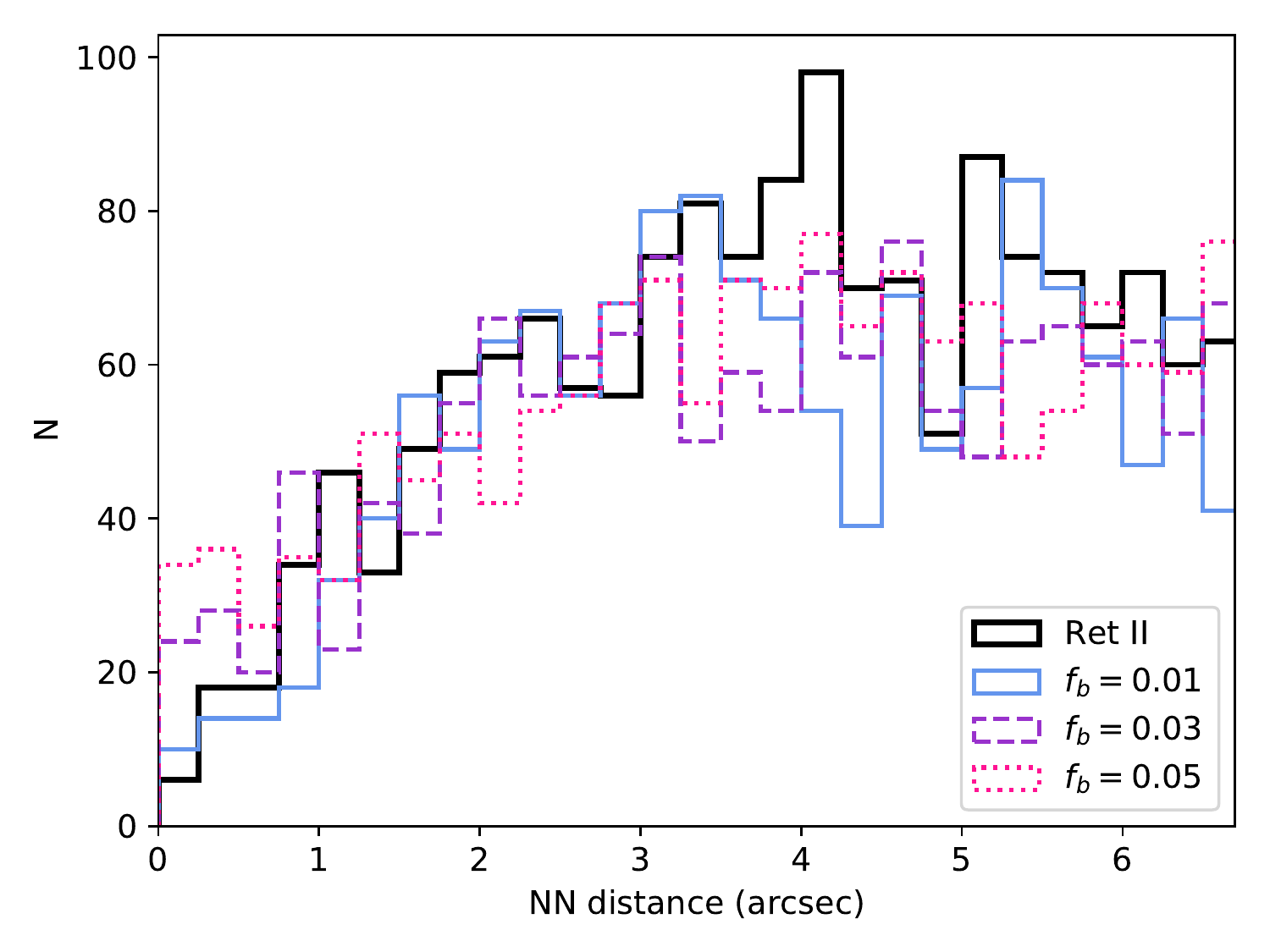}
\caption{\jds{(Left) Comparison}~of the nearest neighbor (NN) distance for Ret~II stars \jds{(black histogram) with mock data sets created by varying the assumed geometry of the galaxy, with no binaries.  The red histogram uses the adopted parameters for Ret~II, and the orange, green, and blue histograms vary the assumed ellipticity or half-light radius.  The mock data generally exhibit fewer pairs of stars at the smallest separations than Ret~II does.  (Right) Comparison of the nearest neighbor (NN) distance for Ret~II stars (black histogram) with mock data sets created by varying the assumed wide binary fraction.  Wide binary fractions of 3\% and 5\% are clearly inconsistent with the data at separations of less than 1\arcsec.}~}
\label{fig_1}
\end{figure*}

In \jds{the left panel of}~Figure \ref{fig_1}, we compare the observed distribution of nearest neighbor (NN) distances for the stars in Ret~II (black histogram) with a simulated population of stars with the same geometry as Ret~II: an exponential profile with a half-light radius of 6\farcm3 and $\epsilon=0.6$ (red histogram).  The construction of this model is described in more detail below (\S~\ref{sec:model}), and in the example shown here the model contains no binary stars.
We notice that the NN distribution of the stars in Ret~II exhibits \jds{a possible}~excess of pairs with small separations compared to what an ellipsoidal reconstruction of Ret~II without binaries yields.  Could this difference be due to the presence of wide binaries?  \jds{In the right panel of Figure~\ref{fig_1}, we compare the Ret~II NN distribution with models with a non-zero binary fraction to show that the data are sensitive to this quantity.}

\section{Methods}\label{sec:method}

\subsection{Constructing a Model Galaxy with Binary Stars}
\label{sec:model}
We construct a 3D ellipsoidal model of a Ret II-like system based on the assumed half-light radius of $R_h=6\farcm3$ and ellipticity of $\epsilon=0.6$.
Assuming an exponential profile for Ret II, we first project the surface density to a 3D radial profile through Abel's integral:
\be
j(r)=-\frac{1}{\pi}\int_r^{\infty}\frac{dR}{\sqrt{R^2-r^2}}\frac{dI}{dR},
\label{eq:surface_dens}
\ee
where $I(R)$ is the surface density profile, modeled as $I(R)\propto e^{-R/R_e}$. Here, $R_e$ is the projected exponential scale radius, which is related to the projected half-light radius by $R_e=R_h/1.68$. 
The radial distances of the stars in the model galaxy are sampled from $r^2 j(r)$, and the projected $x,y$ positions of the stars in an elliptical distribution are given by: 
\be
x=r \sin(\theta) \cos(\phi),~y=(1-e)~r \sin(\theta) \sin(\phi),
\ee
where $\phi$ is sampled uniformly between $(0, 2\pi)$, and $\cos(\theta)$ is sampled uniformly between $(0,1)$.

When including a population of binary stars in the model (with binary fraction $f_b$), the positions of the primary stars are modeled as above, and the positions of the secondary stars are given by:
\be
x_s=x_p+d \cos(\eta),~y_s=y_p+d \cos(\zeta) \sin(\eta),
\ee
where the ``p" and ``s" subscripts refer to the primary and secondary stars in a binary system, respectively, and $d$ is the projected separation sampled from a \jds{power-law}~distribution \jds{$p(d) \propto d^{\beta}$}.
The inclination angle $(\zeta)$ is sampled uniformly in $\cos(\zeta)$ between $(0,1)$, and the binary \mts{phase ($\eta$)}~ is sampled uniformly between $(0,2\pi)$.  

\jds{To match the geometry of the observational data set, we rotate the coordinate system to the position angle of Ret~II and remove stars that fall outside the outline of the HST mosaic displayed in Fig.~\ref{fig:spatial_coverage}.  Both the single stars and binaries that are removed through this step are replaced by an equal number of additional stars (drawn from the distributions described above) located within the HST footprint.}~

\subsection{Assigning Magnitudes to Single and Binary Stars }
\label{sec:modelmags}

\begin{figure}
\hspace{-0.2in}
\centering
\includegraphics[width=\columnwidth]{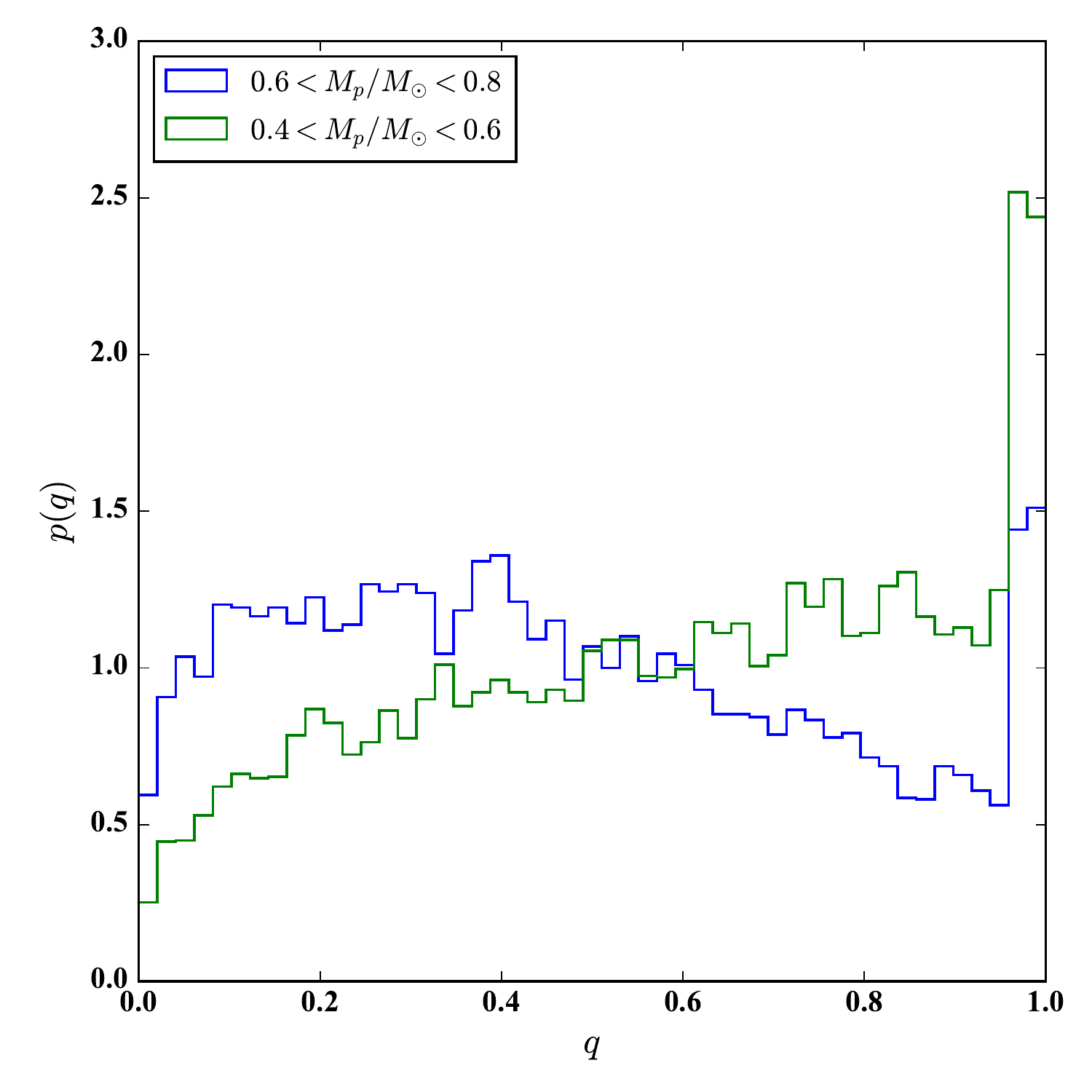}
\caption{Realizations of the distribution of binary mass ratios ($q$) for two separate mass bins of $0.4\le M_p\le0.6~\msun$, and $0.6\le M_p\le0.8~\msun$.  The parameters describing the mass ratio distributions are adopted from \citet{El-Badry2019MNRAS}.
The sharp peak at $q=1$ is due to the excess of twin binaries observed in the solar neighborhood. }
\label{fig_q_dist}
\end{figure}

We assign a mass to each simulated star assuming a \citet{Kroupa:2001ki} initial mass function (IMF).\footnote{We do not expect the assumption of a particular form for the IMF to have a significant impact on the derived binary fraction, since the masses are used primarily to determine whether the model stars fall in the observable magnitude range.  In \S~\ref{sec:imf_model}, we describe our model to determine the IMF directly from the HST data, with results presented in \S~\ref{sec:imf_constraints}.}  For the primary stars, we include stars between the mass limits of 0.2 and 0.8 $\msun$ (lower-mass stars are not detectable in the HST data, while more massive stars are no longer present in Ret~II because of its old age).  For secondary stars, we define their masses to follow the derived mass ratio distribution from the analysis of binaries in the Milky Way \citep{Kouwenhoven2009AA,Chulkov2021MNRAS,El-Badry2019MNRAS}:
\begin{equation}
    \label{eq:q}
p(q) = \dfrac{dN}{dq} \sim \begin{cases} p_{\gamma}(q), ~0\le q<q_{\rm twin}\\
1, ~q_{\rm twin}\le q\le1\\
\end{cases}
\end{equation}
where

\begin{equation}
    p_\gamma\left(q\right)\propto \left(\frac{q}{q_{\rm break}}\right)^{\gamma_{\rm smallq}}\left[1+\left(\frac{q}{q_{\rm break}}\right)^{1/\Delta}\right]^{\left(\gamma_{\rm largeq}-\gamma_{\rm smallq}\right)\Delta}.
    \label{eq:smoothly_broken}
\end{equation}
We adopt $q_{\rm break}=0.5$, and $\Delta=0.3$. The other parameters, $\gamma_{\rm smallq}$, $\gamma_{\rm largeq}$, $q_{\rm twin}$, and $F_{\rm twin}$ depend on the primary mass of the star in the binary and the projected separation of the binaries. We adopt values for these parameters from Table G1 of \citet{El-Badry2019MNRAS}, where there is no evidence for changes as a function of projected separation ($d$) in the range $d > 3000$~AU probed by our data, so we use a single average number for each parameter rather than allowing them to vary. $q_{\rm twin}$ determines the mass ratio above which there is an excess of twin binaries, and $F_{\rm twin}$ is a measure of the relative abundance of the twin binaries to all binaries in a given bin of primary star mass and binary separation. Most relevant for Ret~II are the numbers associated with primary masses between $0.6\le M_p\le0.8~\msun$ and $0.4\le M_p\le0.6~\msun$ for separations between $50\le d/{\rm AU}\le350$.
Figure \ref{fig_q_dist} shows random draws from the mass ratio distribution for the two mass bins we consider. 
 
We determine absolute magnitudes for the simulated stars based on their stellar masses by interpolating the same PARSEC isochrone employed in \S~\ref{sec:data} for member selection.  We convert the absolute magnitudes into apparent magnitudes assuming a distance of 31.4~kpc \citep{MutluPakdil2018ApJ}  and extinction values of $A_{F606W} = 0.045$~mag and $A_{F814W} = 0.028$~mag \citep{Schlafly:2011}.  For binary systems that are unresolved by HST, we combine the magnitudes of the primary and secondary stars.  For resolved binaries where the secondary star is fainter than our magnitude limit, we include only the primary magnitude.

\subsection{Nearest Neighbor Distribution and MCMC Fits}

We then determine the NN separation of each star in the simulated dwarf galaxy.  At the wavelength of the F606W filter, the angular resolution of HST is $\sim0\farcs1$, so we treat binaries with projected NN separation less than $0\farcs1$ as single stars.\footnote{\jds{Consistent with expectations from the HST resolution and the design of DAOPHOT, the closest pair found in the catalog has a separation of 0\farcs17.  Our results do not depend on the precise value assumed for the minimum detectable separation.~}~}~Doing so reduces the number of simulated stars, and to compensate for it we add new single stars to the galaxy so that the total number of simulated stars matches the total number of stars in the Ret II data, which is $n_T=2587$.
 
After computing the projected NN distances for the simulated dwarf galaxy stars, we bin the data into $0\farcs5$-wide bins for separations up to 15\arcsec, with the observed data binned in the same way.

The model presented in \S\ref{sec:model} contains a total of \jds{four}~parameters: the half-light radius and ellipticity of the galaxy, the binary fraction ($f_b$), and the \jds{slope ($\beta$)}~of the \jds{power-law}~distribution of binary separations in AU.  Because the half-light radius is well-determined from the photometric data, and because it would be computationally inefficient to evaluate the integral in Equation~\ref{eq:surface_dens} in each Monte Carlo iteration, we adopt a fixed value for half light radius $R_{h}$, leaving \jds{three}~parameters whose values we wish to determine.
In the Bayesian formalism, the probability of the model parameters 
\jds{$\Lambda = \{f_b, e, \beta\}$}~given the data is proportional to the likelihood of the data. We compare the binned histograms of the model and observed NN distributions by defining a multi-nomial likelihood:
\be
\mathcal{L}(D\mid \Lambda)=\frac{n_T!}{n_1!...n_k!} p_1^{n_1}...p_k^{n_k},
\ee
 where $n_T=\Sigma_{i=1}^{i=k} n_i$ is the total number of stars, and $p_i$ is the probability of each bin, which we obtain from the normalized histogram of the projected NN distribution of each simulation.
$n_i$ represents the number of stars in each bin from the Ret II data.   

We use both the Dynesty \citep{Speagle2020MNRAS} and emcee \citep{emcee2013PASP} Markov Chain Monte Carlo (MCMC) samplers to construct posterior probability distributions for the four parameters.  Initial comparisons with the Ret~II data showed that with a stellar sample of this size, the likelihood function is quite noisy, such that successive fits with the same model can produce noticeably different results.  We therefore draw \jds{10}~random samples according to the model parameters in each MCMC step and evaluate the likelihood function based on the average of those \jds{10}~samples rather than using a single realization of the model.  Relatedly, we observe that a fraction of the MCMC walkers that begin too far from the preferred parameter values converge slowly, if at all.  To prevent these walkers from unnecessarily broadening the posterior probability distributions, we re-initialize the walkers with the lowest 25\% of likelihood values to bring them into agreement with the remainder of the walkers.  We repeat this process three times to ensure convergence.

\subsection{Initial Mass Function}
\label{sec:imf_model}

\begin{figure}
\hspace{-0.2in}
\includegraphics[width=3.75in]{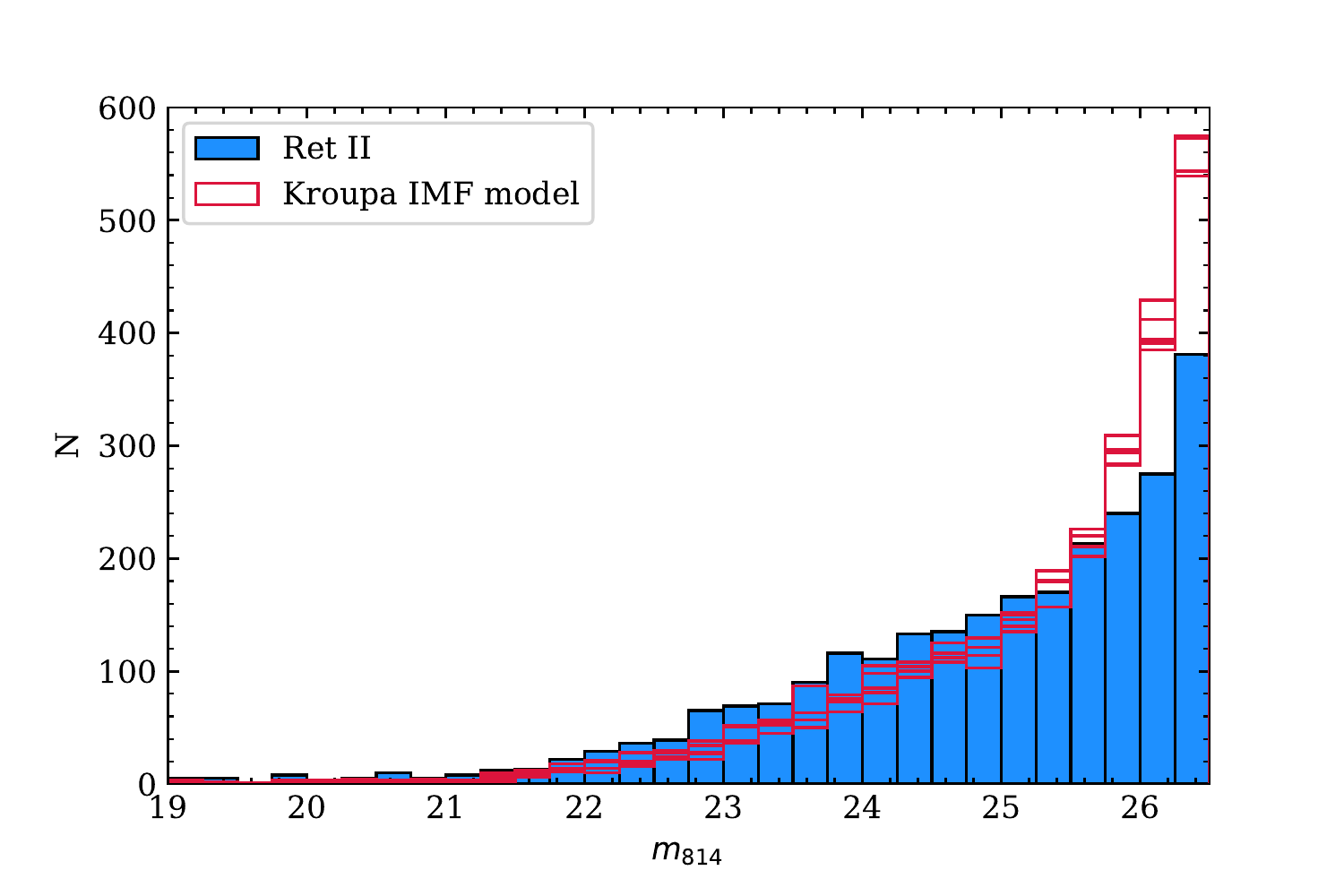}
\caption{Comparison between the observed F814W stellar luminosity function of Ret~II (blue histogram) and a model drawn from a \citet{Kroupa:2001ki} IMF (red).  Five realizations of the Kroupa model are displayed to provide a sense of the scatter associated with the random draws.  Here we have assumed a binary fraction of $f_b = 0.07$ (see \S\ref{sec:fit_to_data}), but different values produce similar results.  The plotted model uses the binary mass ratio distribution for $0.4~\msun < M < 0.6~\msun$, but the results for the $0.6~\msun < M < 0.8~\msun$ mass ratio distribution are similar.  Relative to the data, a Kroupa IMF produces too many faint stars ($m_{814} \gtrsim 25.75$) and too few stars at brighter magnitudes ($m_{814} \lesssim 25$).}
\label{fig:kroupa}
\end{figure}

In addition to producing a predicted NN distribution, the model above also leads to a predicted magnitude distribution.  Comparing this prediction with the observed magnitude distribution of Ret~II can constrain the IMF of the galaxy.  As shown in  Fig.~\ref{fig:kroupa}, the default model with a Kroupa IMF is a poor match to the data for any value of the binary fraction.  We therefore repeat the exercise of determining magnitudes for the model population of stars in \S~\ref{sec:modelmags} with different IMFs, assuming a power-law form:

\be
\xi(m) \propto m^{-\alpha}
\label{eq_imf_functional_form}
\ee

\noindent
over the observed mass range, with the slope $\alpha$ varying between 1.01 and 2.0.  Single stars and the primary stars of binary systems are sampled according to Eq.~\ref{eq_imf_functional_form}, and the masses of the secondary stars are drawn from the distributions given in Eqs.~\ref{eq:q} and \ref{eq:smoothly_broken}.

\begin{figure*}
\hspace{-0.2in}
\centering
\includegraphics[width=\columnwidth]{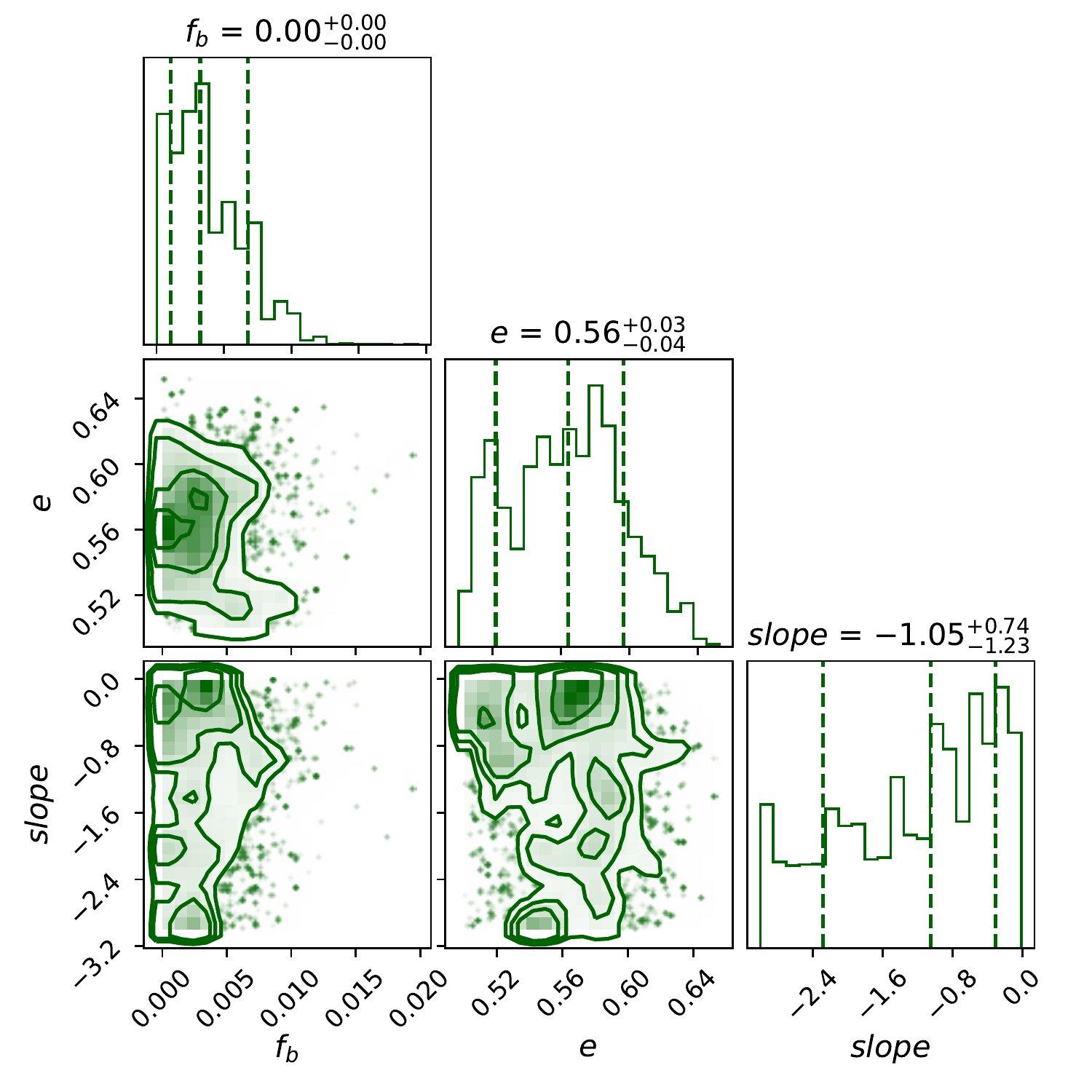}
\includegraphics[width=\columnwidth]{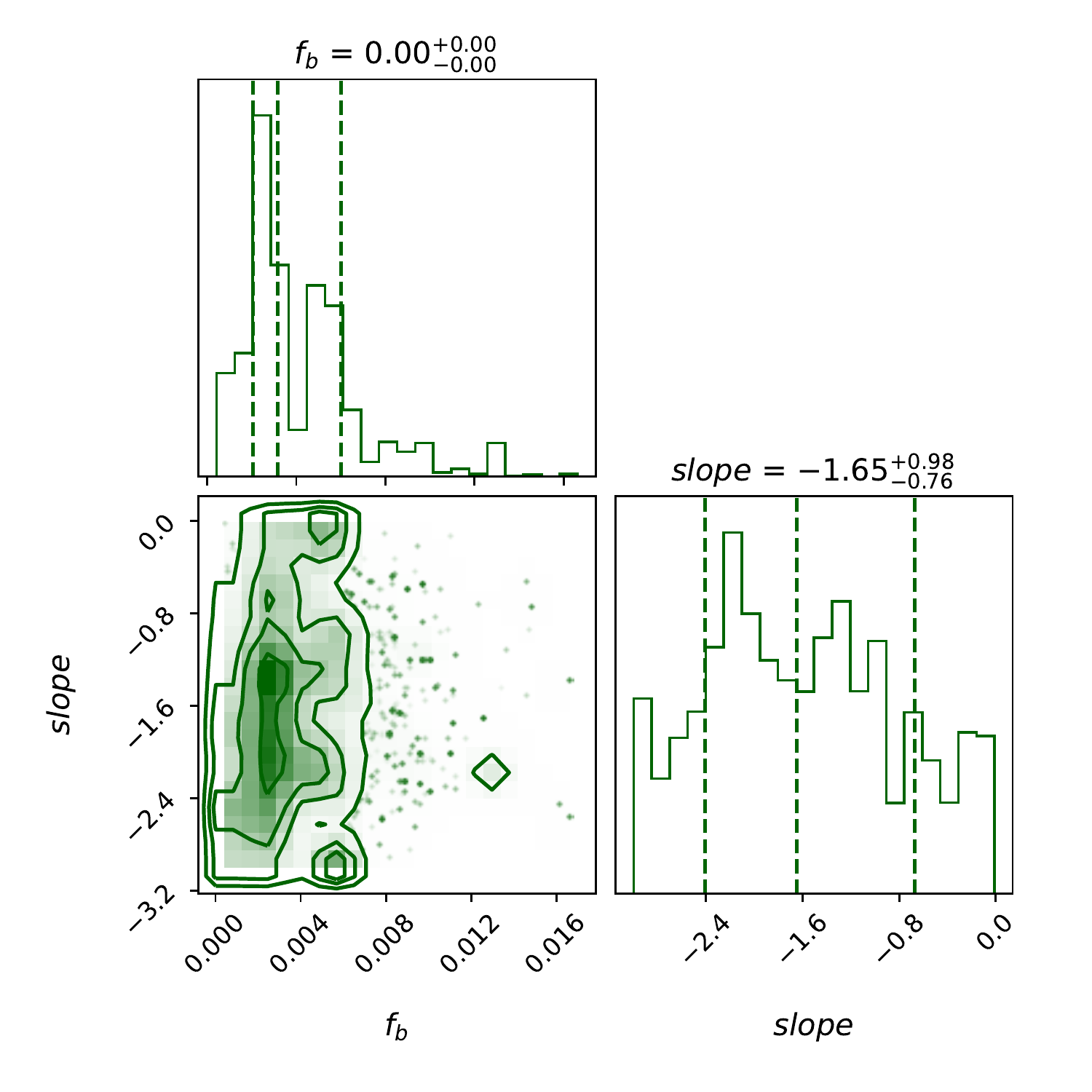}
\caption{\emph{Left panel:} Posterior distributions of \jds{three}~parameters  for a simulated dwarf galaxy with $f_b=0$, $R_h=6\farcm3$, and $e=0.6$. The recovered binary fraction of \jds{$f_b=0.003^{+0.004}_{-0.002}$ and ellipticity of $e=0.56^{+0.04}_{-0.03}$}~are in agreement with the \jds{input}~values. \emph{Right panel:} Posterior distributions of \jds{two}~parameters of our model fitting a simulated dwarf with the same input parameters but the ellipticity held fixed. \jds{The recovered binary fraction is the same, but the uncertainties are smaller than}~when the ellipticity is included as a parameter in the MCMC fit.}
\label{fig_sim_r_h_63_4d}
\end{figure*}

\section{Results} 

\subsection{Tests with Mock Data}

First, we test our method by carrying out MCMC fits on mock data sets (generated as described above) with known parameters.  All mock data are constructed to match the sample size of the Ret~II data set, and all MCMC simulations adopt top-hat priors \jds{for the binary fraction and power-law slope of the separation distribution with boundaries of $0\le f_b\le1$ and $-2.0 \le \beta \le -1.01$, respectively\footnote{\jds{We also explored broader priors for the slope and found that they did not result in any significant differences.~}~}, and a Gaussian prior centered at $\epsilon = 0.6$ with $\sigma = 0.03$ for the ellipticity.}~Since our initial goal is to test the null hypothesis that the binary fraction is zero in Ret~II, the first simulated dwarf contains no binaries ($f_b=0$).  We then run an MCMC simulation to fit for the parameters of the model in this \jds{3D}~parameter space and show the posterior distribution of each parameter in the left panel of Figure \ref{fig_sim_r_h_63_4d}.
The recovered posterior for the binary fraction is \jds{$f_b=0.003^{+0.004}_{-0.002}$}, and the ellipticity is \jds{$e=0.56^{+0.04}_{-0.03}$.}~These values are consistent with the input values for the simulated dwarf, confirming that we can recover a zero binary fraction with confidence.  The median derived \jds{value}~for the \jds{slope of the}~binary separation \jds{distribution is}~also consistent with the true \jds{value,}~but the posterior probability \jds{distribution is}~so broad that the \jds{constraint on this parameter is not}~useful.

We also explore holding the ellipticity of the galaxy fixed in the fit and using the MCMC procedure to determine only the binary parameters, since the ellipticity is already rather tightly constrained by the photometry.  In this case, the recovered binary fraction is unchanged, at \jds{$f_b=0.003^{+0.003}_{-0.001}$, but the uncertainties are somewhat smaller,}~indicating that a \jds{fit with fewer parameters}~may be beneficial in determining more accurate binary properties.  The results of this fit are displayed in the right panel of Figure \ref{fig_sim_r_h_63_4d}.  \jds{Fixing both the ellipticity and the slope of the separation distribution produces very similar results.}~

\mts{Finally, we investigated the impact of assuming an incorrect half-light radius.  We constructed a simulated galaxy with a half-light radius of 5\farcm5, more than 2-$\sigma$ away from the true value, and then fit that data set with $R_h=6\farcm3$ (and vice versa). Our results indicate that assuming a half-light radius offset from the true value by substantial amounts has a negligible effect on the recovered binary fraction.} 

Extending these tests to $f_b > 0$, we find that \jds{the fits generally recover the input binary fraction within the uncertainties regardless of which fit parameters are held fixed.  Thus, although}~in our analysis below we do not know the intrinsic separation distribution for binaries in Ret~II \jds{or the true geometry of the galaxy, we conclude that there is no evidence that our inferred binary fraction should be biased.}

\subsection{Fits to the Observed Ret~II Data Set}
\label{sec:fit_to_data}

\begin{figure*}
\hspace{-0.2in}
\centering
\includegraphics[width=\columnwidth]{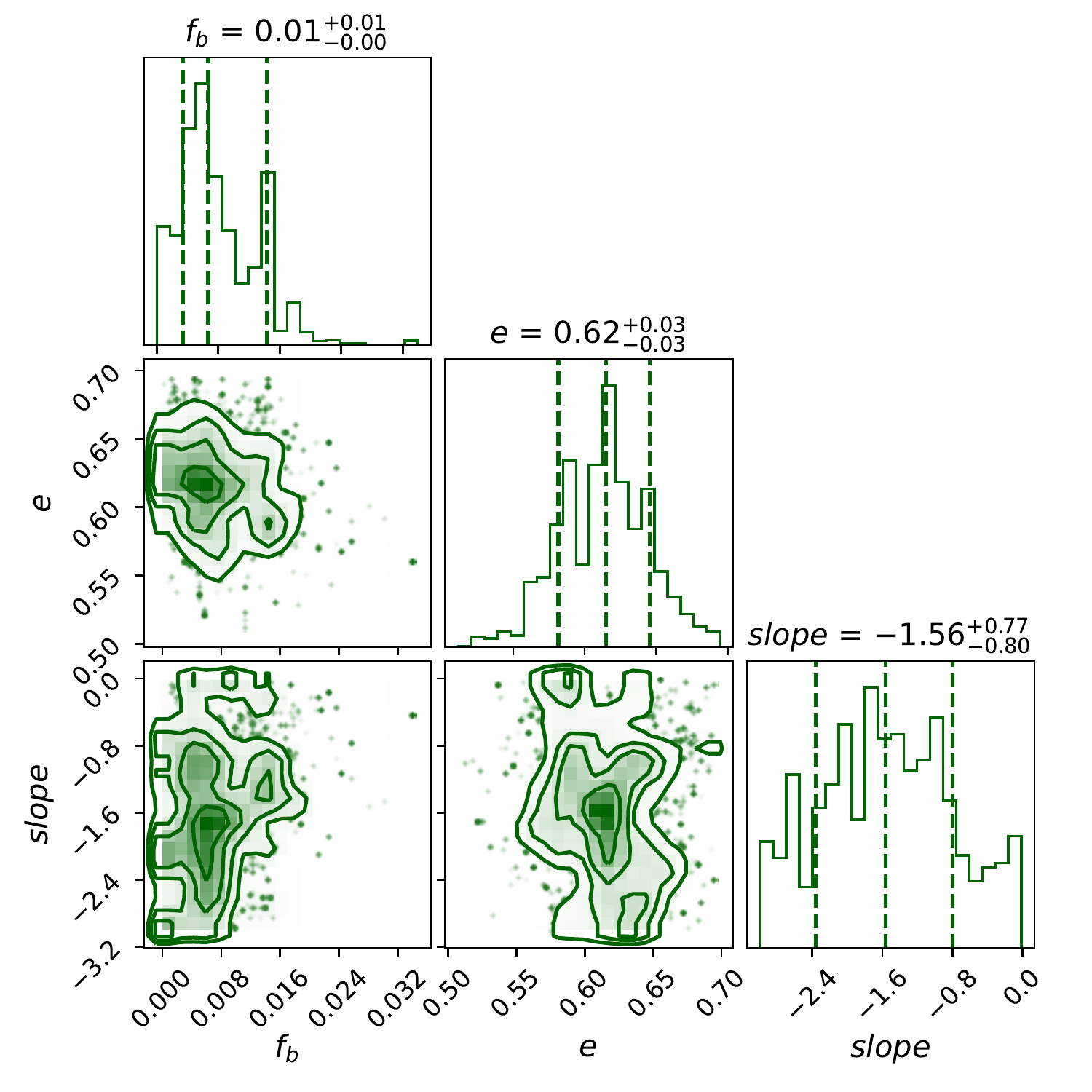}
\includegraphics[width=\columnwidth]{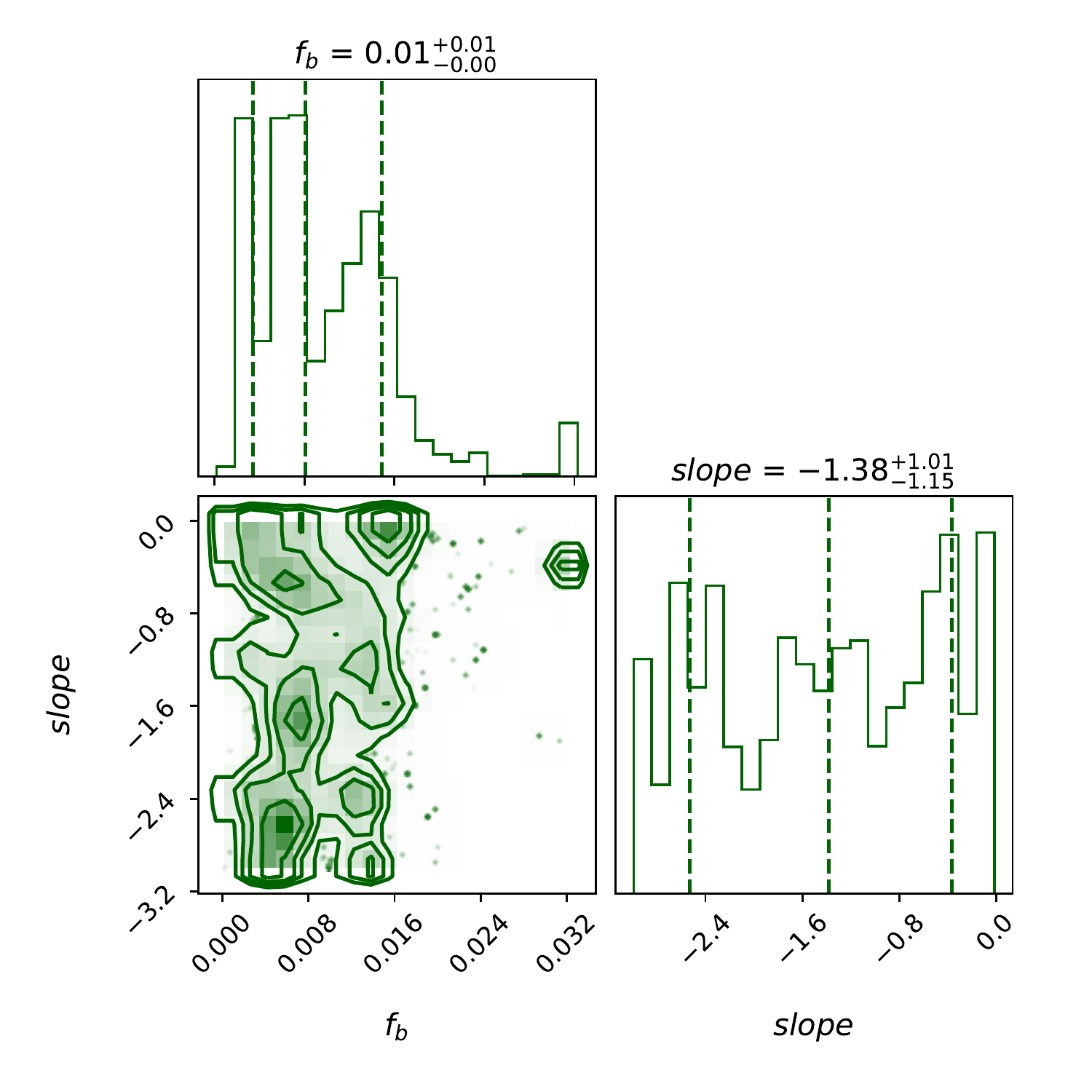}
\caption{\emph{Left panel:} Posterior distributions for our four parameter model fitting the Ret~II nearest neighbor separations, assuming a half-light radius of $R_h=6\farcm3$. \emph{Right panel:} Posterior distributions for our three parameter model fitting Ret~II, assuming $R_h=6\farcm3$ and fixing $e=0.6$. The recovered posterior distribution of $f_b=0.07^{+0.04}_{-0.03}$ is a statistically significant result compared to what we recovered for a simulated dwarf containing no binaries (see Figure \ref{fig_sim_r_h_63_4d}).}
\label{fig_ret_II_r_h_63_4d}
\end{figure*}

Next, we move to performing the same analysis on the HST observations of Ret~II, with results displayed in Figure \ref{fig_ret_II_r_h_63_4d}.  With a \jds{three}-parameter fit, the \jds{fraction of stars with a companion at separations between 0\farcs1 and 10\arcsec\ is $f_b=0.007^{+0.008}_{-0.003}$.  The slope of the power-law describing the binary separation distribution is unconstrained by the data, but is consistent with the value of $\beta = -1.6$ found for binary systems in the solar neighborhood \citep[e.g.,][]{Chaname04,Lepine07,El-BadryRix18,Tian20}}~ 


If we instead impose $e=0.6$ and fit for the remaining \jds{two}~parameters,
we recover \jds{an essentially identical}~posterior distribution of \jds{$f_b=0.007^{+0.007}_{-0.004}$.  Finally, with both the ellipticity and the slope ($\beta = -1.6$) fixed, the derived binary fraction remains unchanged at $f_b=0.008^{+0.007}_{-0.005}$.}~ 

Since numerical simulations have shown that binaries with semi-major axis $\gtrsim1$~pc are prone to disruption by tidal forces \citep{Penarrubia:2016}, we also repeat our MCMC simulations with a maximum angular separation of \jds{5\arcsec~(0.75~pc)}~instead of \jds{10\arcsec}.~  We find that the inferred binary fraction does not depend significantly on the range of separations considered in the fit.  \jds{The only choice in our analysis that has any impact on the fit results is the assumed minimum detectable separation between stars.  Increasing the separation limit from 0\farcs1 to 0\farcs15 or 0\farcs2 raises the binary fraction to $f_b = 0.009^{+0.006}_{-0.004}$ or $f_b = 0.013^{+0.005}_{-0.006}$, respectively.  However, we note that both of those values agree with the result for a minimum separation of 0\farcs1 within the uncertainties, so we do not regard these changes as significant.\footnote{\jds{The DAOPHOT documentation indicates that stars separated by at least one angular resolution element should not affect one another.}~}}~

\subsection{IMF Constraints}
\label{sec:imf_constraints}

\begin{figure*}
\epsscale{1.17}
\plottwo{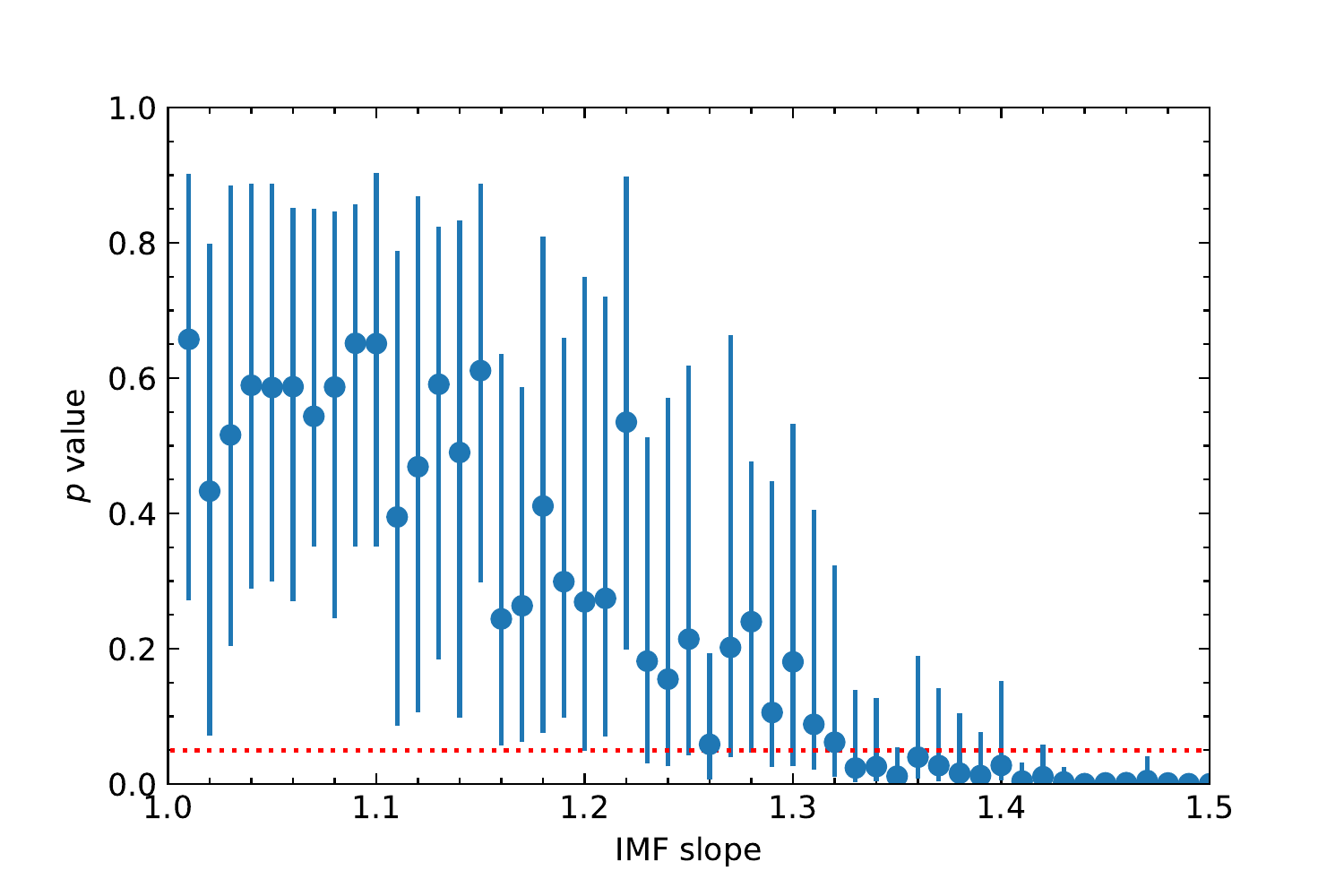}{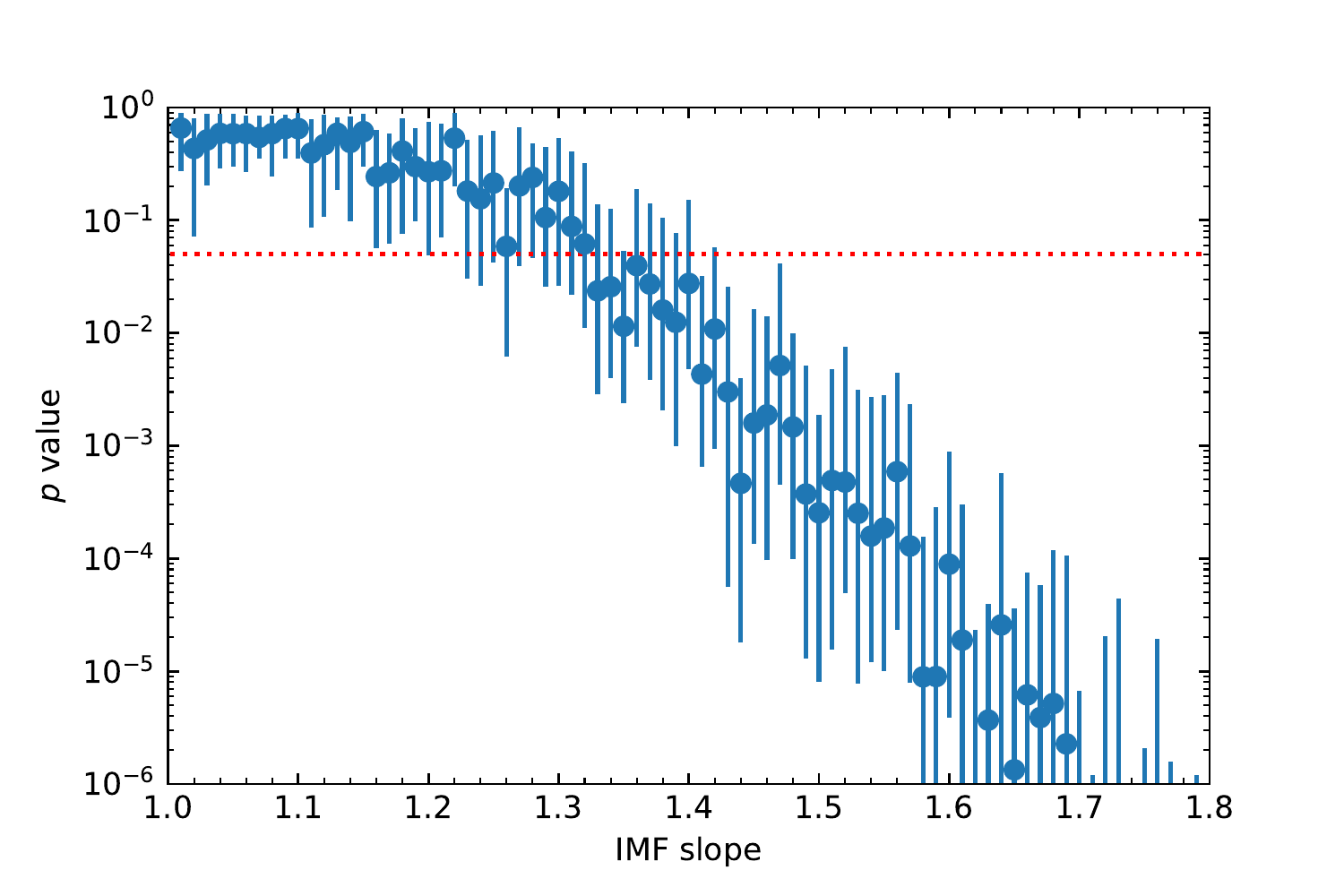}
 \caption{Kolmogorov-Smirnov $p$-value from comparing the observed magnitude distribution in Ret~II with models constructed with different IMF slopes.  The left panel displays the results with a linear scale on the y axis to highlight the good fit obtained at $\alpha \lesssim \jds{1.15}~$, and the right panel uses a logarithmic scale to illustrate the behavior for steeper slopes.  The error bars are plotted as the 16th and 84th percentiles of the distribution of $p$-values obtained for each slope.  The red dotted line shows our limit of $p=0.05$ for a model to be considered a reasonable match to the data.  The distribution of magnitudes in Ret~II strongly prefers a shallow IMF slope of $\alpha \le 1.4$ over the observed mass range of $0.34-0.78~\msun$.}
\label{fig:imf}
\end{figure*}

To assess the IMF in Ret~II quantitatively, we draw random samples of 2587 stellar masses according to IMF power-law indices ranging from $\alpha=1.01$ to $\alpha=2.0$ in steps of 0.01.  \jds{(Here we are assuming that the IMF can be described by a single power law over the mass range probed by our data.)}~We repeat the random draws 100 times for each value of the IMF slope in order to improve the statistics, and convert all stellar masses into observed magnitudes in the F814W band using a PARSEC isochrone.  For each set of model magnitudes, we use a 2-sample Kolmogorov-Smirnov test to calculate the probability $p$ that the observed magnitude distribution and the model are consistent with being drawn from the same probability distribution.  There is some \jds{correlation}~between the \jds{assumed wide}~binary fraction and the IMF slope, but for \jds{the range of wide binary fractions allowed by the data}, the effect on $\alpha$ is small.  We find that the magnitude distribution of stars in Ret~II requires an IMF much shallower than the \citet{salpeter1955} value of $\alpha=2.35$ (see Fig.~\ref{fig:imf}).  For \jds{$f_b = 0.01$,}~the \jds{maximum $p$-values occur at $1.01 \le \alpha \le 1.15$,}~and the 84th percentile of the $p$-values (corresponding to a $1\sigma$ upper limit for a Gaussian distribution) is at least $p=0.05$ for $1.01 \le \alpha \le \jds{1.42}~$.\footnote{As can be seen in Fig.~\ref{fig:imf}, even with 100 $p$-values for each IMF slope, there is some noise remaining in the model comparison.  In particular, although \jds{$\alpha=1.42$}~is the largest value for which we cannot rule out $p \ge 0.05$, the upper limit on $p$ for \jds{$\alpha=1.41$}~is actually slightly less than 0.05.  We assume that these variations are the result of random fluctuations, and ignore them for the purpose of determining the allowed range of IMF slopes.}  Increasing the binary fraction to larger values preserves the preference for $\alpha~\jds{\lesssim}~1.1$ and decreases the quality of the fit for steeper slopes.  The differences between using the $0.4~\msun < M < 0.6~\msun$ and $0.6~\msun < M < 0.8~\msun$ binary mass ratio distributions (as described in \S~\ref{sec:modelmags}) are negligible.  This result is consistent with the IMF slopes determined for the UFDs Canes~Venatici~II, Hercules, Leo~IV, and Ursa~Major~I (all of which are somewhat more luminous than Ret~II) by \citet{Geha2013} and \citet{gennaro2018a}.  \citet{gennaro2018a,gennaro2018b} obtained modestly steeper IMFs for the UFDs Bo{\"o}tes~I and Coma~Berenices, where Coma~Ber is the closest of these galaxies to Ret~II in luminosity \citep{RMunoz2018} and metallicity \citep{Kirby2013ApJ}.  


\section{Discussion and Implications}

We have presented the first observational search for wide binary systems in a UFD via high angular resolution imaging.  Our approach is based on the 2 point correlation function \citep[e.g.,][]{Penarrubia:2016,Kervick:2021}, cast in the form of the nearest neighbor statistic.  We find that the spatial distribution of stars in Ret~II at small separations is consistent with \jds{a fraction of $0.007^{+0.008}_{-0.003}$ of Ret~II stars possessing a companion in the separation range from 0\farcs1 ($3000$~AU) to 10\arcsec\ ($3 \times 10^{5}$~AU).}~

\jds{Combining the Gaia eDR3 wide binary catalog of \citet{El-Badry2021MNRAS} with the Gaia Catalog of Nearby Stars \citep{Smart2021}, we find that in the solar neighborhood, 0.3\% of main sequence stars have a companion at a separation beyond 3000~AU that would be bright enough to observe in our Ret~II data set.  The wide binary fraction in Ret~II is somewhat higher than, but consistent with, this value.  The overall fraction of nearby stars that are in multiple systems is 44\%\ \citep{Raghavan:2010gd}, indicating that for the separation and mass ratio distributions in the Milky Way, only $\sim1$\%\ of binary systems have companions distant enough and bright enough to be detected at a distance of $\sim30$~kpc.  One might therefore suppose that the multiplicity rate in Ret~II is also $\gtrsim50$\%, but the uncertainty on this estimate is quite large.}~

\jds{In recent work, \citet{Hwang21} found that the wide binary fraction in the Milky Way declines by a factor of two from solar metallicity down to $\mathrm{[Fe/H]} = -1$.  If that trend continues to lower metallicities, the wide binary fraction in Ret~II would be expected to be $\sim0.1$\%\ or less, below our measured value (although only modestly inconsistent given the uncertainties).  On the other hand, \citet{El-BadryRix19} derived a flat wide binary fraction with metallicity.  Presuming that the primary difference between the star-forming conditions in Ret~II in the early universe and those prevalent in the progenitors of the Milky Way stellar halo is metallicity, a lack of strong evolution in the wide binary fraction with metallicity is slightly more consistent with our results.  However, more data in both the Milky Way and dwarf galaxies will be needed to draw strong conclusions.}~

Although individual binary stars have been identified spectroscopically in UFDs \citep[e.g.,][]{Frebel2010,Koposov2011,Koch2014,Venn2017,Li2018CarII}, relatively few constraints on the overall binary population in these galaxies are available.  Using hierarchical Bayesian modeling of multi-epoch radial velocity measurements in Segue~1, \citet{Martinez2011} demonstrated that the galaxy must contain either a high fraction of binary stars or shorter mean binary periods than the Milky Way.  However, they were only able to place a weak lower limit on the binary fraction of $f_b \sim 0.05$ at 68\% confidence.  \citet{Minor2019} used a similar method to investigate Ret~II with a smaller spectroscopic sample, concluding that $f_b > 0.5$ at 90\% confidence.  This result \jds{is not obviously inconsistent with our measurements, given the large difference in scales between the binaries we can detect and those whose kinematic signatures are observable.}~

The first photometric determination of the binary fraction in UFDs was by \citet{Geha2013}, who modeled HST color-magnitude diagrams of Hercules and Leo~IV to measure $f_b = 0.47^{+0.16}_{-0.14}$ and $f_b = 0.47^{+0.37}_{-0.17}$, respectively.  More recently, \citet{gennaro2018a} analyzed the entire HST data set of \citet{Brown:2014jn} to study the IMF.  They considered the binary fraction as a nuisance parameter, but found values spanning a very wide range, from $f_b = 0.05^{+0.02}_{-0.05}$ for Canes~Venatici~II to $f_b = 0.61 \pm 0.14$ for Coma~Berenices.  On the other hand, employing deeper near-infrared HST imaging of Com~Ber, \citet{gennaro2018b} determined a smaller binary fraction of $0.25^{+0.08}_{-0.25}$.

Our technique for measuring the binary fraction is distinct from and complementary to the previous methods used in the literature.  Other HST measurements relied on jointly fitting the colors and magnitudes of member stars to determine the IMF and the binary fraction simultaneously based on the photometric differences between single stars and unresolved binaries.  Here, we use stellar positions to statistically identify \emph{spatially resolved} binary stars.  \jds{Encouragingly, the results from different techniques appear to be generally consistent, although as previously mentioned, the uncertainties at this stage are quite large.}~Spectroscopic follow-up of a larger sample of stars in Ret~II and imaging searches for wide binaries in other nearby UFDs \jds{may shed additional light on this subject.}~

\jds{The steepness of the binary separation distribution and the distance of Ret~II provide a relatively narrow range of separations over which wide Ret~II binaries can be identified.  We are therefore unable to determine whether the separation distribution in Ret~II differs in any way from that observed in the Milky Way.}~ 
The spectroscopic studies of Segue~1 and Ret~II, which are sensitive to close binary systems, have suggested a shorter mean period distribution \jds{than in the Milky Way}~\citep{Martinez2011,Minor2019}.  Similarly, \citet{Moe2019ApJ} infer shorter mean periods for metal-poor binary systems in the Milky Way.  More specifically, they find that the close binary fraction is a strong inverse function of metallicity, while the wide binary fraction is independent of metallicity, so the dominant population of close binaries among metal-poor stars shifts the average period to smaller values.  If the same result holds for UFDs, the influence of binary stars on their internal kinematics could be larger than would be estimated by assuming solar neighborhood binary properties.

The IMF in low-mass dwarf galaxies has generally been found to be shallower (more bottom-light) than that in the Milky Way \citep{wyse2002,Geha2013,gennaro2018a,gennaro2018b}.  Our IMF measurement for Ret~II is consistent with these results, with an IMF that is best described by a slope of \jds{$1.01 \le \alpha \le 1.15$}~when fit with a single power-law from $0.34~\msun$ to $0.78~\msun$.  The UFDs therefore remain qualitatively consistent with a picture in which the IMF varies systematically with metallicity, from bottom-heavy IMFs for the most massive galaxies \citep[e.g.,][]{vanDokkumConroy2010,Spiniello2012} to bottom-light IMFs for the least luminous dwarfs (however, see \citealt{gennaro2018a} for possible complications of this model).  If the bottom-light IMF for the surviving low-mass stars in Ret~II can be extrapolated to much higher masses, there are significant implications for the population of supernovae the galaxy hosted at early times, and hence its chemical evolution \citep[e.g.,][]{Jeon2021}.  

The noisiness of the likelihood function and the \jds{large uncertainties}~we obtain on the \jds{fraction of wide binaries}~suggest that the primary limitation of the Ret~II data set is the sample size.  The HST imaging already extends beyond the half-light radius of Ret~II, so imaging over a wider field, as will be possible with, e.g., the Roman Space Telescope \citep{Akeson2019} may only result in modest improvements.  Larger numbers of stars can be obtained from deeper imaging in the near-infrared, either with HST \citep[e.g.,][]{gennaro2018b} or with the James Webb Space Telescope.  Alternatively, it may be preferable to target dwarf galaxies containing larger numbers of stars.  Most such systems are more distant, limiting the spatial resolution that can be obtained, but the recently-discovered UFDs Carina~II and Hydrus~I are each a factor of $\sim2$ more luminous than Ret~II and are located at comparable distances from the Milky Way \jds{\citep{Koposov2018,Torrealba2018}.}~Space-based imaging of these galaxies therefore may provide new insight into the binary star populations in UFDs.  In parallel, spectroscopic observations of larger samples of UFD stars over a longer time baseline ---  the earliest UFD radial velocities are now $\sim15$ years old --- can tighten constraints on close binary systems, and a joint analysis of the spectroscopy and imaging together \jds{could}~improve measurements of the \jds{binary}~separation distribution in these extreme environments.

\acknowledgements 
We gratefully acknowledge the DES collaboration for allowing us to access the unpublished Ret~II data.  We thank Tom Brown and Roberto Avila for providing the photometric catalog used in this paper, as well as for helpful discussions.  \jds{We are grateful to the referee for comments that improved the paper, and to Kareem El-Badry for critical feedback about the results.}~We thank Marla Geha and Ting Li for providing comments on a draft of the paper, and Josh Speagle for advice about MCMC procedures.
This work was supported in part by the Dean’s Competitive Fund for Promising Scholarship at the Faculty of Arts \& Sciences of Harvard University.
This work has made use of data from the European Space Agency (ESA) mission
{\it Gaia} (\url{https://www.cosmos.esa.int/gaia}), processed by the {\it Gaia}
Data Processing and Analysis Consortium (DPAC,
\url{https://www.cosmos.esa.int/web/gaia/dpac/consortium}). Funding for the DPAC
has been provided by national institutions, in particular the institutions
participating in the {\it Gaia} Multilateral Agreement.

\bibliographystyle{yahapj}
\bibliography{ms,jds}
\end{document}